\documentclass[prd,aps,amsmath,amssymb,nofootinbib,preprintnumbers]
{revtex4}

\usepackage{graphicx}
\usepackage{dcolumn}
\usepackage{bm}
\usepackage{amsmath}
\usepackage{amsthm}
\usepackage{amsfonts}
\usepackage{euscript,bbm}
\usepackage{ifthen}
\usepackage{psfrag}
\usepackage{slashed}
\usepackage{hyperref}
\usepackage{color}

\def\ls{\mathrel{\lower4pt\vbox{\lineskip=0pt\baselineskip=0pt
           \hbox{$<$}\hbox{$\sim$}}}}
\def\gs{\mathrel{\lower4pt\vbox{\lineskip=0pt\baselineskip=0pt
           \hbox{$>$}\hbox{$\sim$}}}}
\def\drawbox#1#2{\hrule height#2pt
\hbox{\vrule width#2pt height#1pt \kern#1pt
              \vrule width#2pt}
              \hrule height#2pt}

\def\Asym#1#2{\vcenter{\vbox{\drawbox{#1}{#2}
              \kern-#2pt       
              \drawbox{#1}{#2}}}}

\def\nn{\nonumber}

\newcommand{\be}{\begin{equation}}
\newcommand{\ee}{\end{equation}}
\newcommand{\bea}{\begin{eqnarray}}
\newcommand{\eea}{\end{eqnarray}}

\newcommand{\gsim}{\lower.7ex\hbox{$\;\stackrel{\textstyle>}{\sim}\;$}}
\newcommand{\lsim}{\lower.7ex\hbox{$\;\stackrel{\textstyle<}{\sim}\;$}}

\newcommand{\met}{{E\!\!\!\!/_{\rm T}}}

\newcommand{\sci}[2]{#1$\times$10$^{\text{#2}}$}

\newcommand{\ben}{\begin{enumerate}}
\newcommand{\een}{\end{enumerate}}
\newcommand{\bei}{\begin{itemize}}
\newcommand{\eei}{\end{itemize}}

\begin{document}

\title{Light Higgsino Decays as a Probe of the NMSSM}

\author{Bhaskar Dutta$^{1}$}
\author{Yu Gao$^{1}$}
\author{Bibhushan Shakya$^{2}$}

\affiliation{
$^{1}$~Department of Physics and Astronomy, Mitchell Institute for Fundamental Physics and Astronomy, Texas A\&M University, College Station, TX 77843-4242\\
$^{2}$~Michigan Center for Theoretical Physics, University of Michigan, Ann Arbor, MI 48109, USA
}

\begin{abstract}
In the Next-to-Minimal Supersymmetric Standard Model (NMSSM), a sizable coupling $\lambda$  between the singlet and Higgs fields can naturally accommodate the observed Higgs boson mass of 125 GeV. This large coupling also results in a large separation between the Higgsino and singlino mass scales in the neutralino sector to evade stringent constraints from direct detection experiments. Most of the natural parameter space in this setup therefore contains light Higgsinos that can decay into the singlino and the 125 GeV Higgs. If the Higgsinos are to be light enough to be produced at the LHC, the mass gap $m_{\tilde\chi^0_2} -(m_{\tilde\chi^0_1} +m_h)$ is generally fairly low, resulting in small missing energy. We study the collider phenomenology of this process and demonstrate that the 4$b+j+\met$ signal arising from $pp\rightarrow\tilde\chi_i^0\tilde\chi_j^0jj$ process can be a viable channel to search for this low mass gap region at the 14 TeV LHC. In addition, we find that a potential signal from the LHC search, in tandem with direct detection experiments, can distinguish the NMSSM from the analogous process in the MSSM, where the bino plays the role of the singlino, for positive values of the $\mu$ parameter even when the additional singlet-dominated (pseudo)scalars are absent or indistinguishable from the Higgs in the decay.

\end{abstract}
\noindent MIFPA-14-35 \\MCTP-14-42\hspace{0.2cm}

\maketitle

\section{Introduction}

The search for supersymmetry in the form of colored superpartners at the Large Hadron Collider (LHC) has so far yielded null results. The exclusion limits on squark ($\tilde{q}$) and gluino ($\tilde{g}$) masses, when they are comparable, are approximately $1.5$ TeV at $95\%$ CL with $20$ fb$^{-1}$ of integrated luminosity \cite{:2012rz, Aad:2012hm, :2012mfa, LHCsquarkgluino20ifb}. The LHC has, however, observed a new boson consistent with the Standard Model (SM) Higgs, with a mass of $125$ GeV \cite{LHCHiggs}, which can be used to obtain insight on the possible underlying model of supersymmetry and promising directions for future searches. In the MSSM, $m_h\approx 125$ GeV requires multi-TeV stops and hence suffers from sub-percent level fine-tuning. In contrast, the next-to-minimal extension of the MSSM (NMSSM), which extends the MSSM by a singlet superfield, can realize this mass very naturally if the coupling $\lambda$ between the singlet and Higgs fields is sufficiently large. 
 
A direct test of the singlet's existence at colliders is via possible modification of the cross-sections of various Higgs production processes. This can be difficult  if the SM Higgs mixing with the singlet field is small. Another approach is to measure  the weak decay of a new particle in the theory into the scalar or pseudoscalar, for instance the decay of $\tilde\chi^0_2$ into  $b\bar{b}$ or $\tau^\pm$ resonance states with invariant masses different from  the 125 GeV Higgs. Such scenarios with additional (pseudo)scalars much lighter than $\chi^0_2$ and their phenomenological searches have been previously studied in~\cite{bib:lightNMSSM}. If a light (pseudo)scalar below 125 GeV is discovered in the future, the NMSSM extension will be the preferred interpretation over a low $M_A$ MSSM scenario as the MSSM scalar couples to the SM particles and faces severe constraints from both collider Higgs search bounds and direct detection. 

In this paper, we focus on regions of parameter space where the singlet (pseudo)scalar is heavier or in close vicinity of the 125 GeV Higgs, i.e. scenarios without a distinguishable second scalar, so that the above signatures are absent, and the sfermions and gluino are sufficiently heavy to evade detection. One must then seek other avenues to probe the supersymmetric sector.    

As with the MSSM, the most natural regions of the NMSSM are characterized by a low $\mu$ parameter close to the weak scale, corresponding to light Higgsinos. As will be explained in a later section, the combination of large $\lambda$, necessary to naturally obtain a 125 GeV Higgs, and direct detection constraints result in a singlino sufficiently lighter than the Higgsinos, with a mass gap generally larger than 125 GeV. A large portion of the most natural surviving parameter space is therefore characterized by Higgsinos decaying into a singlino and an on-shell Higgs.

The purpose of this paper is to explore the prospects of studying this process $\tilde\chi^0_{2,3}\rightarrow \tilde\chi^0_1+Z, h$ at the LHC, extracting information about the underlying model and distinguishing it from the analogous process in the MSSM (where the bino can play the role of the singlino) using details of the process and correlations with direct detection results. Section~\ref{sect:benchmark} describes the NMSSM model and the relevant parameter space, and lists the benchmark points used in our study. Section~\ref{sect:results} examines the direct production of the $\tilde\chi$'s in association with two additional jets, which are helpful in identifying the signal when the mass gap between the neutralinos is not much greater than the Higgs mass, at the LHC. Section~\ref{sect:pheno} discusses correlations with direct detection searches and features of the collider signal that can distinguish this NMSSM signal from the analogous MSSM signal. We summarize and conclude our study in Section~\ref{sect:conclusion}.

\section{Model and Benchmark scenarios}
\label{sect:benchmark}

The NMSSM adds a singlet superfield to the MSSM matter content, contributing the following terms to the superpotential~\cite{Ellwanger:2009dp}
\be 
W_{\text{Higgs}} \supset \lambda \hat{S} \hat{H}_u\cdot \hat{H}_d 
+\frac{\kappa}{3}\hat{S}^3.
\label{eq:superpotential}
\ee
When the singlet acquires a vacuum expectation value (vev), the first term generates an effective $\mu$ parameter. The singlet superfield adds three physical fields to the particle content: a CP-even scalar, a CP-odd scalar, and a neutral fermion, the singlino; these will all play crucial roles in our analysis. 

The soft SUSY-breaking Lagrangian contains the following new terms:
\be
m_S^2|S|^2+(\lambda A_\lambda H_u\cdot H_d S+\frac{1}{3}\kappa A_\kappa S^3+~{\rm h.c.})\,,
\ee

The SM Higgs-like scalar mass at one loop order is given, for small $A_\lambda,~A_\kappa$, by~\cite{Ellwanger:2009dp}
\bea 
M^2_{h} &\approx & M_Z^2 \cos^2 2\beta +\lambda^2 v^2 \sin^2 2\beta -\frac{\lambda^2}{\kappa^2}v^2(\lambda -\kappa \sin 2\beta)^2 + \nn \\
 && + \frac{3m_t^4}{4\pi^2v^2}\left( \ln \left(\frac{m_T^2}{m_t^2}\right)
 +\frac{(A_t-\mu \cot\beta)^2}{m_T^2}\left(1-\frac{(A_t-\mu \cot\beta)^2}{12m_T^2}\right)\right) 
 \label{eq:hmass}
\eea
where $m_T$ is the stop mass and $A_t$ is the top trilinear coupling in the soft SUSY-breaking terms. The first and final terms are the tree and loop level contributions familiar from the MSSM. The second term is the additional tree level contribution in the NMSSM, coming from the first term in Eq.\,\ref{eq:superpotential}; for moderate values of tan$\beta$ and large $\lambda\gsim 0.6$, this can easily account for the observed 125 GeV mass for the Higgs even when the stop loop corrections are small. The third term denotes the contribution from mixing with the singlet scalar when it is heavier than the SM-like Higgs. 

While a large value of $\lambda$ is preferred, it is constrained to $\lambda \lsim0.7$ in the NMSSM if the theory is to remain perturbative up to the GUT scale. The allowed range can be extended to $\lambda\lsim2$ with a Landau pole below the GUT scale but above 10 TeV and therefore consistent with precision electroweak tests, leading to a framework dubbed `$\lambda$-SUSY' \cite{Barbieri:2006bg}. Note that such large values of $\lambda$ cause the tree level Higgs mass to overshoot the 125 GeV mark, and the mixing with the singlet is then required to lower the mass down to 125 GeV, as a consequence of which the SM-like Higgs has some non-negligible singlet content \cite{singletmixing}.

\begin{figure}[h]
\includegraphics[height=5cm]{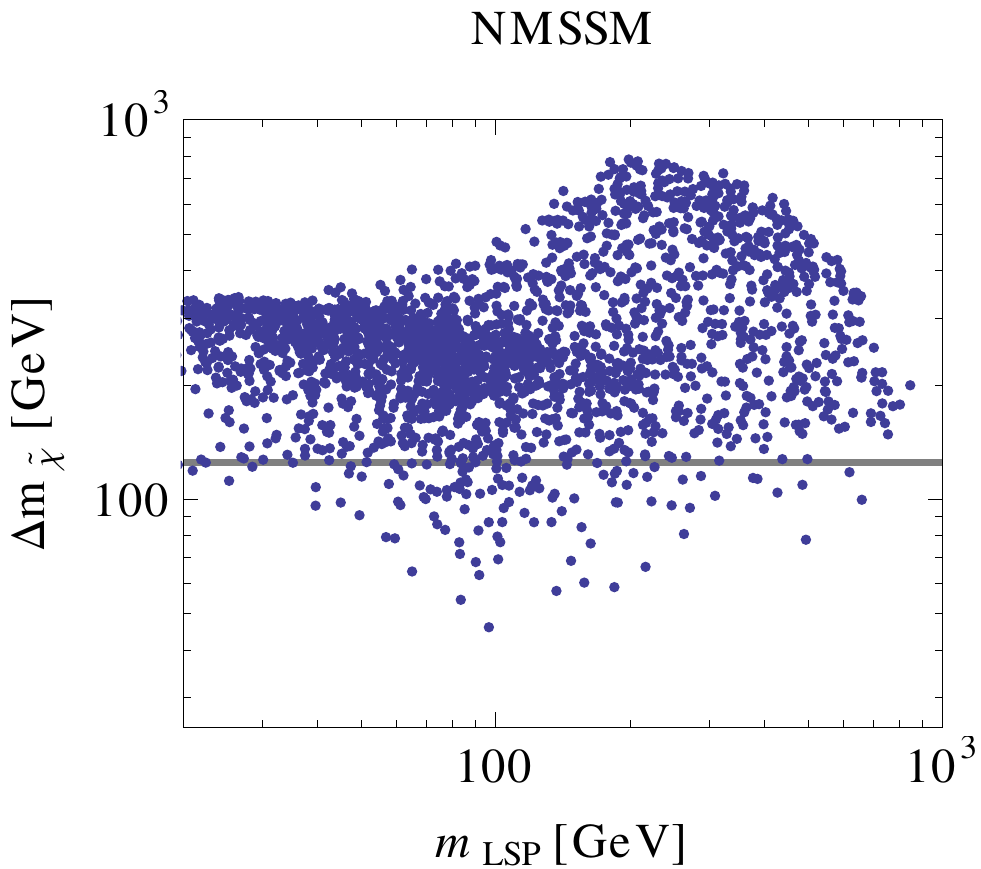}
\caption{Neutralino mass splitting required to evade direct detection bounds from LUX \cite{lux}, here taken to be a constant $5\times 10^{-10}$pb, as a function of LSP mass. The splitting is greater than $m_h=125$GeV (horizontal line) in most of the surviving parameter space.}
\label{fig:splitting}
\end{figure}

In the neutralino sector, the bino and wino masses are unrelated to the Higgs sector, so we assume them to be heavy and thus decoupled. As is well known, naturalness of the electroweak scale requires the $\mu$ parameter to be of the same order, so the Higgsinos are expected to be at the weak scale. The singlino mass is also tied to the $\mu$ parameter, hence the following Higgsino-singlino block of the neutralino mass matrix is relevant:
\be 
\begin{array}{c}
{\scriptsize \tilde{H}_d} \\
{\scriptsize \tilde{H}_u}\\
{\scriptsize \tilde{S}}
\end{array}
\left(
\begin{array}{ccc}
0 & -\mu  & -v\lambda \sin\beta \\
-\mu  & 0 & -v\lambda \cos\beta \\
-v\lambda \sin\beta& -v\lambda \cos\beta & \frac{2\kappa}{\lambda}\mu
\end{array}
\right)
\ee
A large $\lambda$ required for the 125 GeV Higgs has several implications in the neutralino sector.  First, the singlino is generally lighter than the Higgsinos, i.e. $2\kappa\,\textless\, \lambda$. Second, since the off-diagonal term is proportional to $\lambda$, this implies a large mixing between the singlino and Higgsinos. Third, if the Lightest Supersymmetric Particle (LSP) is a Higgsino-singlino mixture, this implies a large direct detection cross section with nuclei, which is in tension with direct detection bounds \cite{lux}; evading the constraints requires the LSP to be a fairly pure singlino unless accidental cancellations are in effect\,\cite{ddtuning2}. This can be accomplished by raising the Higgsino mass scale relative to the singlino mass scale\,\footnote{Another alternative, which we ignore in this paper, is to raise the singlino mass above $\mu$ to obtain a pure Higgsino LSP. This is both difficult to accomplish with a large $\lambda$ and leads to collider signals that have been studied elsewhere in literature \cite{Delannoy:2013ata}.}. A scan over the required splitting between the two lightest (Higgsino- and singlino-like) mass eigenstates necessary to avoid the current LUX bound \cite{lux} when scattering occurs through a SM-like Higgs only is plotted in Fig.\,\ref{fig:splitting}, and shows that the splitting can be larger than $m_h=125$ GeV (horizontal line) in most of the parameter space, enabling the decay $\tilde\chi^0_{2,3}\rightarrow \tilde\chi^0_1+h$.

To study this decay process at the LHC, we focus on a few benchmark points; these are listed in Table~\ref{tab:benchmarks}\,\,\footnote{The software package {\it NMSSMTools}\cite{bib:nmssmtools} is used to evaluate the particle spectrum and mixing angles. Due to the sfermion masses being set at a multi-TeV scale, the masses and mixings in Table~\ref{tab:benchmarks} are also evaluated after RGE running to the multi-TeV scale. One can expect corrections up to ${\cal O}(10^{-1})$ factors to a neutralino mass directly obtained from the input parameter values at the weak scale. Two-loop corrections to the scalar masses are included since its impact on the scalar mixing angles can be significant~\cite{Degrassi:2009yq}.}. We restrict ourselves to the difficult scenario where both the second scalar and the lightest psuedoscalar are either heavier than or near-degenerate (within $\sim 10\%$ in mass) with the observed Higgs so that they are either kinematically inaccessible or not easily distinguished even if produced in the decay process. To enable meaningful comparisons across the points, we have chosen similar numerical values for the parameters. Large $\lambda$ and small tan$\beta$ are chosen to facilitate the 125 GeV Higgs mass naturally. Note that there is in general substantial mixing with the singlet, as denoted by large values of $S_{h3}^2$; current data allows up to $50\%$ mixing \cite{singletmixing}, and we choose points that saturate this bound, although the exact fraction does not play a crucial role in our conclusions. $\mu$ is chosen to have the heavier neutralinos at $\sim 270$ GeV in order to allow significant electroweak production at the LHC. The value of $\kappa$ is then chosen to get the LSP at $\sim 140$ GeV, so that the NLSP-LSP mass splitting is slightly greater than the Higgs mass. The mass hierarchy essentially fixes the singlino component $N_{15}^2$ in the LSP. The $\sigma_{SI}$ and $\tilde{\xi}^{Zh}$ values will be relevant in Section\,\ref{sect:pheno}.




\begin{table}[h]
\begin{tabular}{c|cccccc|cc|c|c|ccc|c}
\hline
Benchmark &\ $\lambda$\ &\ $\kappa$\ &\ $\mu$\ &\ $\tan\beta$\ &\ $A_\lambda$\ &\ $A_\kappa$\ & $N_{15}^2$ & $S_{h3}^2$ & $m_{a_1}$  &$\sigma_{\text{SI}}$ (pb) & $m_{\tilde\chi^0_1}$ & $m_{\tilde\chi^0_2}$ & $m_{\tilde\chi^0_3}$& $\tilde{\xi}^{Zh}$\\
 \hline
A &0.8 & 0.25 &220 & 2.9 & 710 & 45 & 62\% & 50\% & 161&\sci{9}{-11}&143&270 &270 &2.1\\
\hline
B &0.8 & 0.24 &210 & 2.9 & 682 & 100 & 62\% & 42\% & 115&\sci{1.6}{-10}&133 &259 &261&0.7\\
\hline
C&0.8 & 0.25 &230 & 2.9 & 710 & 100 & 64\% & 25\% & 119 &\sci{3.4}{-10} & 150 & 279 & 279&0.7\\
\hline\hline
A' {\scriptsize~(MSSM), $M_1=140$GeV} & - & - & 260 & 20 & - & - & 93\%{\scriptsize ($\tilde{B}$)} &- & $10^3$ & \sci{2.3}{-9} &134 &270 &275&1.6\\
\hline
\end{tabular}
\caption{Benchmark NMSSM points used in this study. Point A' is a MSSM counterpart to Point A with a similar light neutralino spectrum as described in the text. $\tilde{\xi}^{Zh}$ is the ratio of decay into Z to decay into h (see \ref{sec:collidersignal}).
}
\label{tab:benchmarks}
\end{table}

The benchmark points are almost identical, with differences chosen to highlight important aspects of the decay process. 

\begin{itemize}
\item Benchmark Point A is a typical low $\mu_{}$ NMSSM scenario where $\tilde\chi_2^0$, $\tilde\chi_3^0$ only decay into the LSP and $s_1$ or $Z$, and $s_1$, the lightest singlet, is a significant mixture between the SM-like Higgs and the singlet. 
\item Benchmark Point B is similar to Point A, but the lightest pseudoscalar $a_1$ is also light and appears in the decay process, taking up a sizable branching fraction due to its large singlet component.
\item Benchmark Point C is similar to Point B, except the lightest (125 GeV) scalar has a smaller singlet fraction.
\item Benchmark Point A' is an MSSM point that mimics the neutralino mass spectrum of Point A with the bino in place of the singlino as the LSP; the wino remains decoupled. 
\end{itemize}



\begin{figure}[h]
\includegraphics[height=5cm]{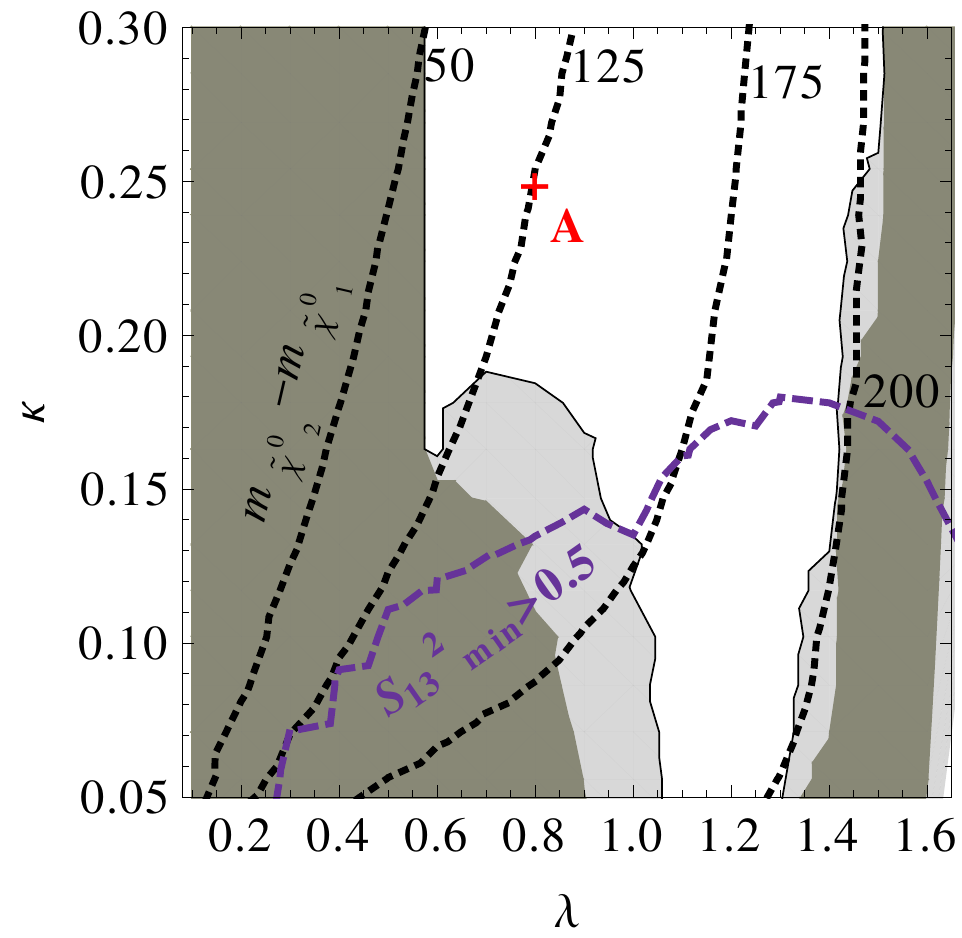}
\includegraphics[height=5cm]{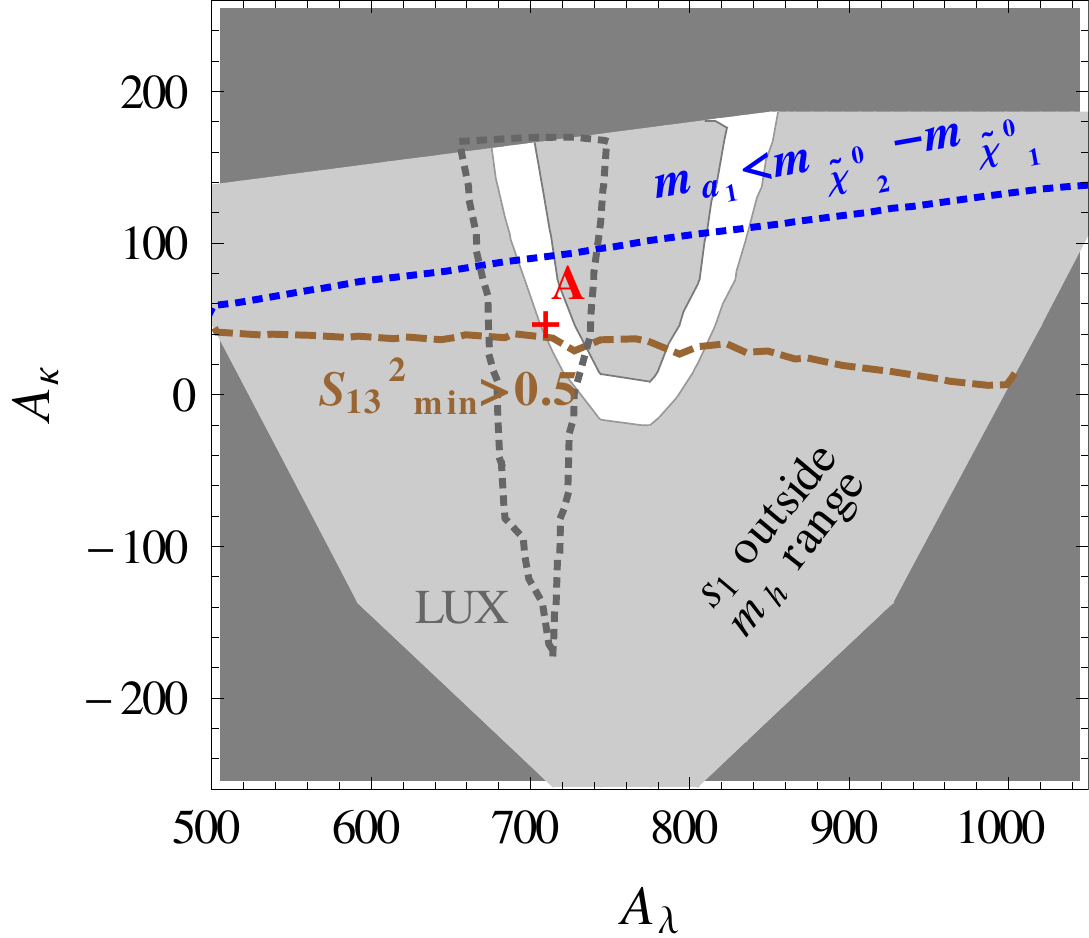}
\caption{($\lambda, \kappa$) and ($A_\lambda, A_\kappa$) planes near the Benchmark Point A.} 
\label{fig:higgsmass}
\end{figure}

To visualize the characteristics of the benchmark points, Fig.~\ref{fig:higgsmass} illustrates a slice of the NMSSM parameter space around Point A. 
At lower $\tan\beta$ the mass gap between $\tilde\chi^0_2$, $\tilde\chi^0_1$ shrinks significantly, while at larger $\tan\beta$ it  becomes more difficult to keep all scalar masses positive. In the left panel of Fig.~\ref{fig:higgsmass}, the dotted lines show contours of constant $m_{\tilde\chi^0_2}-m_{\tilde\chi^0_1}$. In both figures, the region above the dashed curve corresponds to a minimal singlet mixing below 50\% (marginalized over $A_{\lambda}$ and $A_{\kappa}$) in the observed Higgs. In the left figure, the regions with $m_{s_1}$ outside  the allowed 122.7-128.7 GeV range are shaded in dark gray and the region where ${a_1}$ is always lighter than ${s_1}$ is shaded in light gray. In order to have both minimal $S_{13}^2$ and  $m_{s_1}$ in the observed Higgs mass range,  $A_\lambda, A_\kappa$ are varied over all  possible values (hence point A lies away from the $S_{13}^2= 50\%$ boundary in the figure). The right panel shows the \{$A_\lambda, A_\kappa$\} parameter space, where the rest of the parameters are fixed to those of Point A. The white half-circle band on the right figure shows the region where ${s_1}$ occurs in the allowed range of the Higgs mass, while the regions above the brown and blue curves denote the regions where $S_{13}^2< 50\%$   and the light pseudoscalar $a_1$ may show up in the $\tilde\chi^0_2$ decays respectively. The latest direct detection bound from LUX \cite{lux} is also plotted in the right figure; the parameter space inside the triangular dotted boundary is allowed by the LUX bound. 

Point B resides very close to Point A in the NMSSM parameter space, and would cross the $m_{a_1}<m_{\tilde\chi^0_2}-m_{\tilde\chi^0_1}$ line in its $\{A_\lambda, A_\kappa\}$ subspace. Point B also contains a pseudoscalar near-degenerate with the Higgs that can emerge in the neutralino decay (but this point is potentially distinguishable from Point A, as will be discussed later). Notice that Points A and B have relatively large singlet fractions $S^2_{h3}$ in $s_1$ but are still marginally consistent with the LHC Higgs measurements. However, this large singlet fraction in $s_1$ is not necessary in general, as seen at Point C, which lies nearby in parameter space and has a 125 GeV Higgs candidate with a much smaller singlet component. 

The following section discusses the search strategy and prospects for the production and subsequent decay process $\tilde\chi^0_{2,3}\rightarrow \tilde\chi^0_1+Z, h$. This will be followed by a discussion in the subsequent section of how this process can reveal information about the underlying model.

\section{Prospects at the 14 TeV LHC}
\label{sect:results}

For relatively light $\tilde\chi^0_2, \tilde\chi^0_3$ accessible at the LHC, the mass difference $m_{\tilde\chi^0_2}-(m_{\tilde\chi^0_1}+m_h)$ is generally expected to be small. This mass gap is $\leq 5$\,GeV for the benchmark points listed in the previous section. 
Such a small mass gap region would possess very little missing energy ($\met$) unless the $\tilde\chi^0$ pair is boosted against additional jets. This is analogous to a stop decay ($\tilde t\rightarrow t+\tilde\chi^0_1$) or sbottom decay  ($\tilde b\rightarrow b+\tilde\chi^0_1$) process where $m_{\tilde t(b)}-(m_{\tilde\chi^0_1}+m_{t(b)})$ is very small~\cite{stop}. A similar scenario also occurs with a Higgsino LSP, where the mass difference between $\tilde\chi^0_2(\tilde\chi^{\pm}_1)$ and $\tilde\chi^0_1$ is very small~\cite{xerxes}.
Inspired by the fact that a pair of additional jets in the vector boson fusion topology~\cite{stop, Delannoy:2013ata} can transversely boost the invisible system to generate $\met$, we adopt the $pp\rightarrow\tilde\chi_i^0\tilde\chi^0_jjj$ process as a viable search channel that can allow a large missing energy cut to be made. It is interesting to note that $pp\rightarrow\tilde\chi_i^0\tilde\chi_j^0j$ process, although at one lower order of parton level radiation, will not include the vector boson fusion diagrams and can be less effective than $pp\rightarrow\tilde\chi_i^0\tilde\chi_j^0jj$ at our benchmark points.
The two tagging jets along with large missing energy can significantly suppress SM backgrounds.




In order to beat the SM $t\bar{t}$+jets background, we require that both $Z$ and $s_1$ decay into $b\bar{b}$, thus forming a $4b$ final state with no leptons but significant $\met$. The invariant mass of final state $b\bar{b}$ resonances can then be reconstructed for $Z$ and $h$. To utilize the $t\bar{t}$ analyses in Ref.~\cite{Dutta:2012kx}, we adopt the following kinematic cuts: 
\begin{enumerate}
\item number of jets $N_j>4$ and missing energy $\met > 150$ GeV,
\item lepton and hadronic $\tau$ veto in the central region $|\eta|<2.5$,
\item the leading central jet $P_T>100$ GeV,
\item four tagged $b$ jets with $|\eta|<2.5$. 
\end{enumerate}
For $b$-tagging, we assume a generic 70\% efficiency for any $b$-jet candidate with $P_T>30$ GeV. The probability of tagging all 4 $b$ jets in a signal event is thus $0.7^4=28\%$.

The resulting cross-sections with these cuts for the NMSSM benchmark points from Section \ref{sect:benchmark} are listed in Table~\ref{tab:results}.  We used the NMSSM package~\cite{bib:nmssm_package} for {\it MadGraph 5}~\cite{Alwall:2011uj} to calculate the $\tilde\chi^0\tilde\chi^0+jj$ process, then used {\it Pythia}~\cite{Sjostrand:2006za} as part of the {\it Pythia-PGS} interface~\cite{bib:pythia-pgs} to fully decay and shower the final state. A jet algorithm with cone radius 0.4 was then applied to group the partons into jets, and the $b$-jets within $|\eta|<2.5$ were considered for tagging candidates. The surviving signal cross-sections after the cuts are at the $0.01-0.1\,$fb level.

\begin{table}[h]
\begin{tabular}{c|c|c|c}
\hline
Cut/probability  & \ \ Point A\ \ & \ \ Point B\   & \ \ Point C\   \\ 
\hline
dijet+$\met$  cuts& 3.5 fb & 3.0 fb &2.6 fb \\
\hline
4 $b$ branching with tagging efficiency &  {0.59\%} & 1.2\% & 1.4\%\\
\hline
$b\bar{b}$ rates  & 0.04 fb &0.07 fb &0.07fb\\
\hline
\end{tabular}
\caption{Kinematic efficiencies and $b\bar{b}$ rates at the NMSSM benchmark points.}
\label{tab:results}
\end{table}


Note that  the $s_1\rightarrow b\bar{b}$ branching ratio depends on the $H_u$,$H_d$,$S$ mixing in $s_1$, which can vary at different parameter points. For Point A The final event cross-section is $2\times 10^{-2}$ fb and $b\bar{b}$ resonances at either $Z$ or the `Higgs' mass can be reconstructed with 0.04 fb. While the production cross-section is similar at Point B,  the $\tilde\chi^0_{2,3}\rightarrow a_1+ \tilde\chi^0_1$  decay mode allows the NLSPs to have a greater branching probability into (pseudo)scalars, hence a greater probability of finding a 4$b$ final state; at Point B, the branching ratio into $b\bar{b}$ for $s_1$ and $a_1$ are 41\% and 91\%  respectively. Point C is similar to Point B for collider phenomenology, which suggests that the singlet fraction in the SM-like Higgs is not extremely crucial for our study. 

Next, we briefly discuss the major SM backgrounds. For the leading SM $t\bar{t}$ background, the pre-cut $pp\rightarrow t\bar{t}$ (inclusive) cross-section is $\sim$500 pb at 14 TeV. The kinematic cuts (1-3) reduce the $t\bar{t}$ background by a factor of
${\cal O}({10^{-4}})$~\cite{Dutta:2012kx}. Since $t\bar{t}$ only produces two $b$ jets, two additional $b$ mis-tags must occur to fake a signal event, which, assuming a 1\% fake rate, is expected at a probability of $(10^{-1})^2$. Improvements in the $b$ fake rate to 0.5\%~\cite{Anderson:2013kxz} can further reduce this background. A $b\bar{b}$ resonance mass window cut around Higgs and Z masses can provide an additional reduction by a factor of $\sim0.1$~\cite{Khachatryan:2014mma}. Thus the combined efficiency off all these cuts is $10^{-7} - 10^{-6}$ and can bring the SM $t\bar{t}$ background down to a level similar to the signal strength. 

The SM $t\bar{t}h$ background does not require $b$ mis-tags to fake the signal.  However, this process occurs with a smaller pre-cut cross-section of 0.5 pb, and the kinematic cuts (1-3), expected to have ${\cal O}(10^{-4})$ efficiency, suppress this background to less than 0.1\,fb.

The weak processes $pp\rightarrow hh,hZ,ZZ$ + (1-2) jets can also be a background of concern. They can be controlled by the $\met$ cut in combination with the 4$b$ requirement, which select the fully hadronic $h,Z$ decays that yield little $\met$. Their combined pre-cut cross-section is 80\,fb, and the $\met>150$ GeV cut along with the four $b$ jets requirement reduces this background with ${\cal O}(10^{-4})$ efficiency, resulting in a cross-section similar to or less than the signal strength.

Finally, a few comments are in order. The production cross-sections for the listed benchmark points can be considered to be conservative, as the $\tilde\chi^0_2, \tilde\chi^0_3$ are singlino-Higgsino admixtures and a smaller singlino component can enhance the production rate. However, the electroweak production cross-section is fairly small for heavier neutralino masses, hence we do not expect the reach to go beyond $m_{\tilde\chi^0_{2}}\sim400$ GeV. In addition to the neutralinos, charginos can be produced analogously, but their decays do not directly yield $Z$ or $h$ that we base our study on, hence we ignore chargino states in our study. In addition to the hadronic final states studied here, one can also use the leptonic Z decays to study $bblljj+\met$ and  $lllljj+\met$ final states as mentioned in~\cite{Khachatryan:2014mma}; a detailed analysis for all these final states will be presented in a forthcoming study.
   
\section{Distinguishing the NMSSM from the MSSM}
\label{sect:pheno}

The analysis in the previous section suggests that the $\tilde\chi^0_{2,3}\rightarrow \tilde\chi^0_1+Z, h$ process can be observed at the 14 TeV LHC. However, it does not necessarily have to come from the NMSSM. As seen from Benchmark Point A' in Table\,\ref{tab:benchmarks}, the MSSM can mimic the mass spectrum of the first three neutralinos exactly, with the bino in place of the singlino. A bino LSP can also occur in the NMSSM, but in most of the NMSSM parameter space with large $\lambda$ the singlino cannot be decoupled from this process. Hence we ignore this part of the NMSSM parameter space, and in this paper we take a bino-like LSP to correspond to the MSSM, to be contrasted with the NMSSM setup of a singlino-like LSP. In this section we discuss whether complementary information from direct detection experiments and details of the observed signal can distinguish the NMSSM from the MSSM. 

\subsection{Direct Detection Constraints}
\label{sec:directdetection} 
 
 In the absence of light sfermions, dark matter-nuclei scattering is mediated by the CP-even Higgses. In the NMSSM, the relevant coupling between neutralinos i, j and scalar k is given by 
\be
\label{eq:hcNMSSM}
\frac{\lambda}{\sqrt{2}} N_{i5}\,(N_{j4}S_{k1} + N_{j3}S_{k2})+\left(\frac{\lambda}{\sqrt{2}}N_{i3}N_{j4} -\frac{\kappa}{\sqrt{2}}N_{i5}N_{j5}\right)S_{k3} +\,(i\leftrightarrow j)  
\ee
where we have dropped the gaugino contributions as they are taken to be decoupled, and
\bea
\tilde{\chi}_{0i}&=&N_{i3}\tilde{H}_d+N_{i4}\tilde{H}_u+N_{i5}\tilde{S},\nonumber\\
h_k&=&S_{i1}H_d+S_{i2}H_u+S_{i3}S.
\eea

The analogous coupling in the MSSM is
\be 
\frac{g_1}{2} N_{i1}(N_{j4} S_{k2} +N_{j3}S_{k1})+\, (i\leftrightarrow j)  
\label{eq:hcMSSM}
\ee
where we assume a light Bino plays the role of singlino ($N_{i1}$ denotes the bino component of neutralino $i$) and dropped the wino contribution. 

Note that the MSSM coupling is analogous to the first term in the NMSSM coupling in Eq.\,\ref{eq:hcNMSSM}, with $g_1/\sqrt{2}$ instead of $\lambda$, and $N_{j3}$ and $N_{j4}$ interchanged. The term in the second parenthesis in Eq.\,\ref{eq:hcNMSSM} is the contribution from the singlet component of the scalar. For large $\lambda\gsim0.6$, the NMSSM coupling appears to be significantly larger. However, mixing in the Higgs sector, in particular with the singlet, can lead to both a suppression by the singlet fraction of the 125 GeV Higgs as well as accidental cancellations in the various terms in Eq.\,\ref{eq:hcNMSSM} and between contributions from different Higgs mass eigenstates, resulting in a lowered direct detection cross section. This is fairly generic in the NMSSM. As is well known, an analogous cancellation can also occur in the MSSM, but only for negative values of $\mu$\,\cite{Feng:2010ef}. 

\subsection{Collider Signal}
\label{sec:collidersignal}

Since the $\tilde\chi^0_2$ is heavy enough to decay into $\tilde\chi^0_1, h$, the decay into $\tilde\chi^0_1,Z$ is also kinematically allowed. The $\tilde\chi^0_1\tilde\chi^0_2 Z$ coupling vanishes when $\tilde\chi^0_1$ is a pure singlino since the Z only couples to the Higgsino-like fraction of neutralinos. However, even in this scenario, the coupling to the Higgs field results in the production of the longitudinal component of the Z boson, so that the decay into $\tilde\chi^0_1,Z$ still occurs. It is therefore of interest to compare the branching ratios into these states. The weighted final-state neutralino-to-$Z/h$ ratio can be another useful observable in addition to  the production cross-section of the process; we denote it as
\be 
\xi^{Zh} = \left.\sum_{i>1}{BR_{\tilde\chi_i^0}BR(\tilde\chi_i^0\rightarrow\tilde\chi_1^0 Z)}\right /\sum_{i>1}{BR_{\tilde\chi_i^0}BR(\tilde\chi_i^0\rightarrow\tilde\chi_1^0 h^*)},
\label{eq:etaZh}
\ee
where $BR{\tilde\chi^0_i}$ is the number fraction of the produced $\tilde\chi^0_i$ among the final state particles, and $BR(\tilde\chi^0_i\rightarrow\tilde\chi_1^0 X)$ is the decay branching of $\tilde\chi^0_i\rightarrow \tilde\chi_1^0~Z,h^*$. Here $h^*$ can include all non-SM (pseduo)scalars close enough in mass to the 125 GeV Higgs to be indistinguishable from it in the decay process. In the NMSSM, $h^*$ can include  $s_1$ and/or $a_1$, as in our benchmark points B and C; 
in the MSSM, this is strongly constrained by collider and direct detection bounds, so that $h^*$ generally only includes the SM-like Higgs. The subscript $i$ iterates over the neutralinos other than the LSP. 

 When $\tilde\chi^0_2,\tilde\chi^0_3$ are domiantly Higgsino and nearly degenerate, as is the case in our study, $BR_{\tilde\chi^0_2}\approx BR_{\tilde\chi^0_3}$ and Eq.~\ref{eq:etaZh} can be simplified as
\be 
\tilde{\xi}^{Zh} \equiv \frac{
{BR(\tilde\chi_2^0\rightarrow\tilde\chi_1^0 Z)+BR(\tilde\chi_3^0\rightarrow\tilde\chi_1^0 Z)}}{ {BR(\tilde\chi_2^0\rightarrow\tilde\chi_1^0 h^*)+BR(\tilde\chi_3^0\rightarrow\tilde\chi_1^0 h^*)}},
\hspace{0.21cm}
\label{eq:tilded_ratio}
\ee
We use this $\tilde{\xi}^{Zh}$ ratio for our analysis in the next subsection.

\subsection{Correlating Direct Detection and Collider Signals}
\label{sec:colliderDDsignal}

Let us now investigate whether the singlino-Higgsino NMSSM setup can be distinguished from the bino-Higgsino MSSM setup when the additional (pseudo)scalars are absent or indistinguishable from the 125 GeV Higgs in the decay process. 

\begin{figure}[h]
\includegraphics[height=5cm]{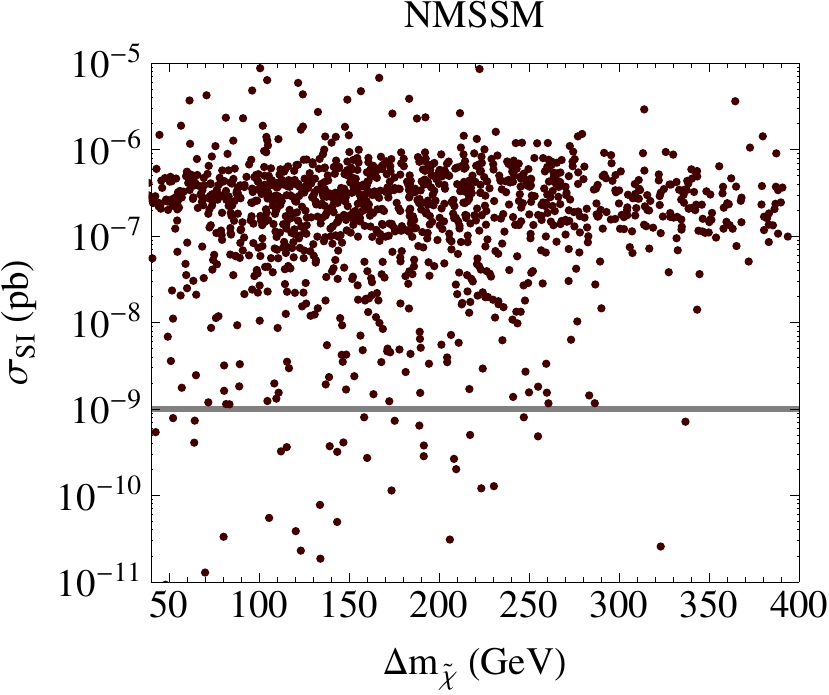}
\includegraphics[height=5cm]{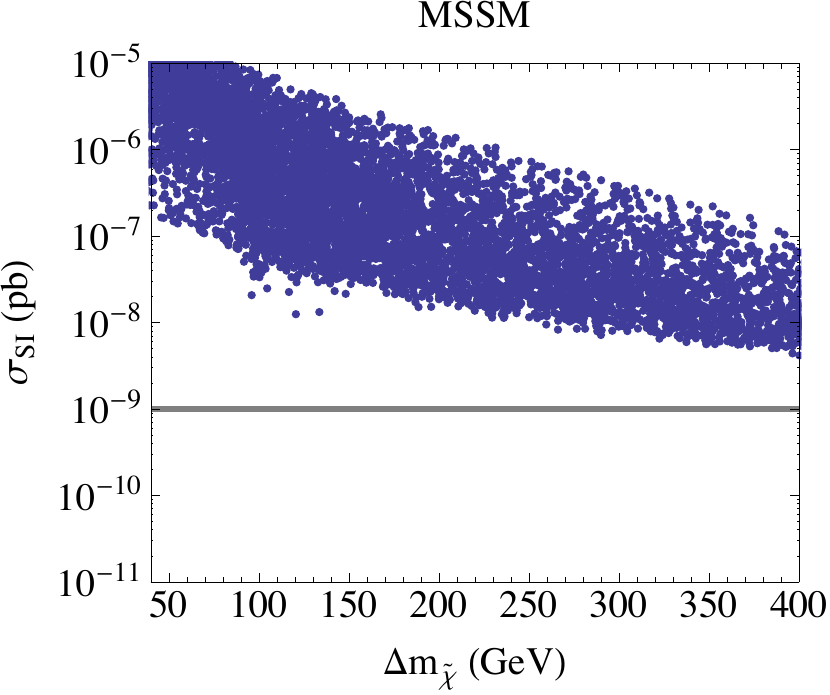}
\caption{Spin-independent direct detection cross section $\sigma_{SI}$ as a function of NLSP-LSP mass splitting in the NMSSM (left) and MSSM (right) for $\mu\,\textless\, 500$ GeV. Current bounds from LUX \cite{lux} rule out cross sections above $\sim10^{-9}$\,pb (horizontal line) in this mass range.} 
\label{fig:directdetection}
\end{figure}

In Sec.\,\ref{sec:directdetection} we discussed the possibility of low direct detection cross-sections in the NMSSM due to numerical cancellations, which can only occur in the MSSM for negative $\mu$.   This feature is clearly visible in the results of a scan plotted in Fig.\,\ref{fig:directdetection}. The scan is restricted to Higgsinos lighter than 500 GeV. Recall that current bounds from LUX \cite{lux} rule out cross sections below $\sim10^{-9}$\,pb in this mass range, denoted by the horizontal line in the plots (the exact bound is mass-dependent, and can be relaxed by up to a factor of 5 in this mass range). The NMSSM parameter space (left) is mostly clustered around a cross section between $10^{-7}$ and $10^{-6}$ pb, but contains points well below this bound, due to the aforementioned numerical cancellations. In contrast, the MSSM scan (right), performed for $\mu\,\textgreater \,0$, shows that the direct detection cross-sections consistently exceed the LUX bound. The cross-sections listed for the benchmark points in Table\,\ref{tab:benchmarks} are in agreement with these observations. 

\begin{figure}[h]
\includegraphics[height=5cm]{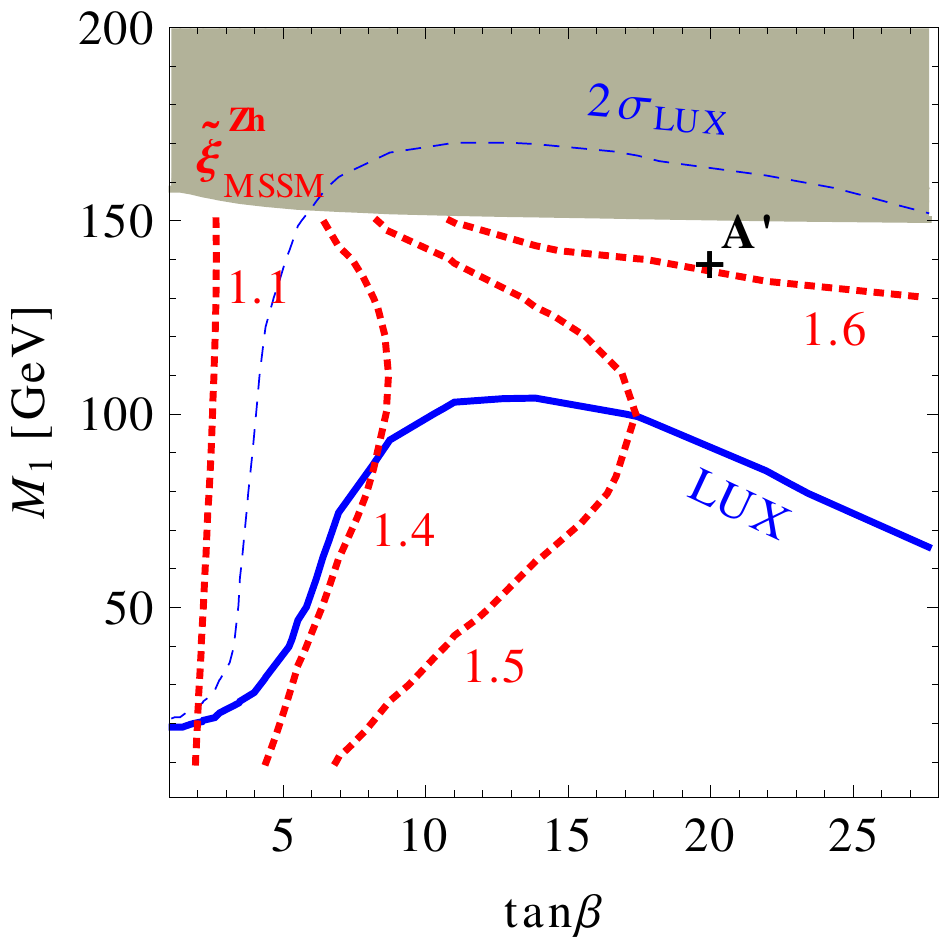}
\caption{LUX bound (blue) and $\tilde{\xi}^{Zh}$ contours (red) in the MSSM for $\mu=260$ GeV. The shaded region denotes $m_{\tilde\chi^0_2}-m_{\tilde\chi^0_1} < m_h$. The LUX bound is significant: the MSSM point A' that mimics the NMSSM benchmark point A (see Table~\ref{tab:benchmarks}), is barely allowed if the LUX bound relaxes by a factor of 2 (denoted by the $2\sigma_{LUX}$ contour).}
\label{fig:mssm_LUX}
\end{figure}

Fig.~\ref{fig:mssm_LUX} shows the interplay between the LUX bound and the bino mass $M_1$ in the MSSM for $\mu=260$ GeV. The LUX bound essentially rules out $M_1\textgreater 100$ GeV, and Benchmark Point A'  is only allowed if the LUX bound is relaxed by a factor of 2 (denoted by the $2\sigma_{LUX}$ contour). 
In contrast, note that Benchmark Point A, corresponding to a similar mass spectrum in the NMSSM, evades the LUX bound by an order of magnitude. 
The current LUX bound allows smaller values of $M_1\leq 50$ GeV, as the LSP becomes a purer bino. 
This region of wider mass gap $m_{\tilde\chi^0_2}-(m_{\tilde\chi^0_1}+m_{h/Z})$ can be probed by Drell-Yan production at the LHC. Likewise, a similarly low $\tilde\chi^0_1$ mass in the NMSSM requires a small $\kappa$, which generally also leads to a light singlet-dominant (pseudo)scalar that can be probed with $\tau^+\tau^-$ or $b\bar{b}$ resonance searches.

These observations therefore point to meaningful correlations between the mass splitting between the two lightest neutralinos and the direct detection cross-section if the LSP is the dark matter. The observation of the decay signal discussed in the previous section at the LHC would favor an NMSSM interpretation, even in the absence of additional direct evidence of the singlet, due to the MSSM being incompatible with bounds from direct detection. We reiterate that this correlation can be avoided in the MSSM with $\mu\,\textless\,0$. 

In addition, details of the decay process, in the form of $\tilde{\xi}^{Zh}$, can also carry useful information. If $s_1$ is the only visible scalar in $\tilde\chi^0_2, \tilde\chi^0_3$ decays in the NMSSM, 
we find that $\tilde{\xi}^{Zh}$ typically occurs in the range $1<\tilde{\xi}^{Zh}<2$ as long as the decay into $s_1$ is not on the brink of kinematic suppression. The same range of $\tilde{\xi}^{Zh}$ also appears in the bino-Higgsino system in the MSSM, as can be seen in Fig.~\ref{fig:mssm_LUX}. This is hardly surprising and can be understood as arising from the correlation between the $h$ and (longitudinal) $Z$ couplings due to the Goldstone equivalence theorem~\cite{Jung:2014bda}.  

\begin{figure}[h]
\includegraphics[height=5cm]{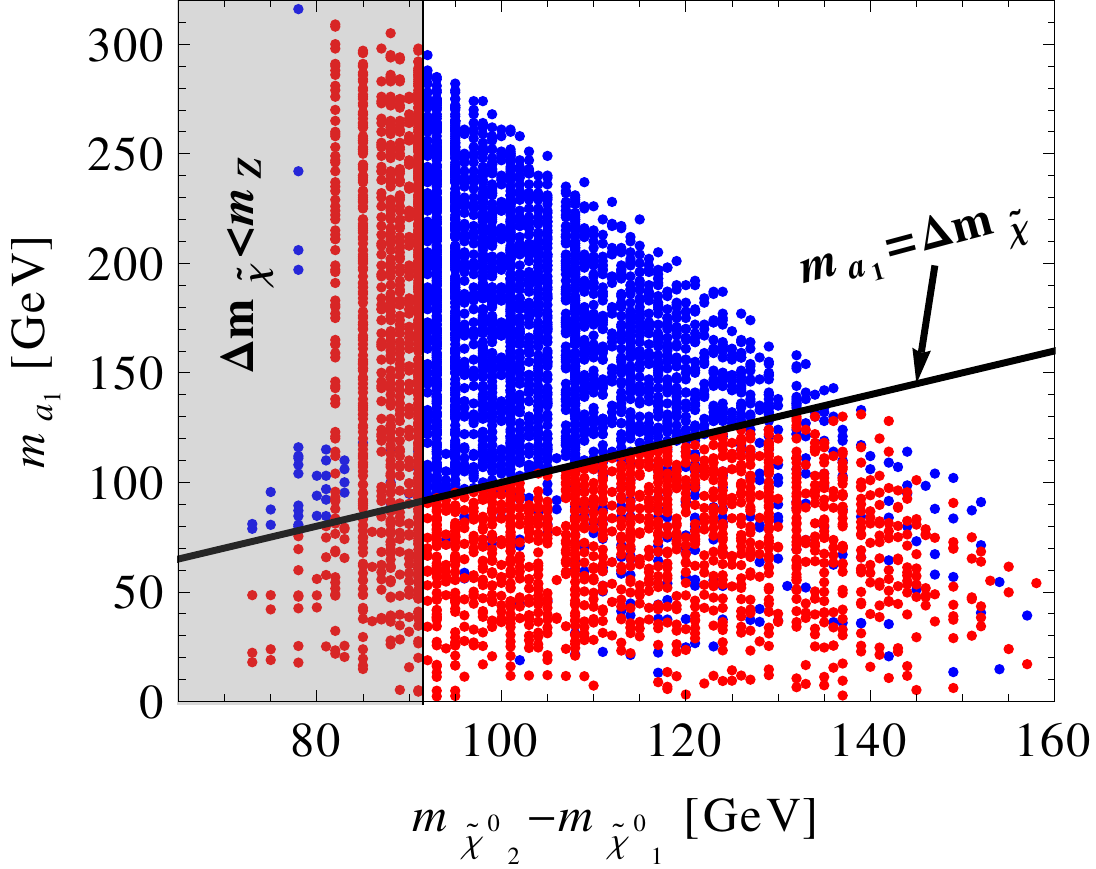}
\caption{Dependence of $\tilde{\xi}^{Zh}$ on whether the pseudoscalar can be produced on shell in the $\tilde{\chi}_{2}^0$ decays. The ratio is greater (less) than 1 for the blue (red) points. In the shaded region, $m_{\tilde\chi^0_2}-m_{\tilde\chi^0_1}\,\textless\,m_Z$, and our 4$b$ on Z/h resonance signal vanishes in the absence of two body NLSP decays.
}
\label{fig:nmssm_ratio}
\end{figure}

There are, however, exceptions to this pattern where the NMSSM clearly deviates from this range. When the decay of $\tilde\chi^0_{2,3}$ into the singlet pseudoscalar $a_1$ is allowed, we find that the $\tilde\chi_{2,3}^0\rightarrow \tilde\chi^0_1 a_1$ decay channel can lead to  $\tilde{\xi}^{Zh}<1$. This can occur if the $a_1$ mass is close enough to 125 GeV that the two decays are indistinguishable at the LHC, as in our benchmark points B and C, which are both characterized by $\tilde{\xi}^{Zh}<1$ (see Table \ref{tab:benchmarks}). Fig.\,\ref{fig:nmssm_ratio} shows that this behavior is very generic; $m_a\,\textless\, m_{\tilde\chi^0_2}-m_{\tilde\chi^0_1}$ is generally (but not always) correlated with $\tilde{\xi}^{Zh}<1$, which are denoted by red points on the plot. Since the singlet field, unlike the SU(2) doublet, does not contribute any component to the Z boson, a kinematically accessible $a_1$ contributes to the decay to $h^*$ without contributing anything to the decay to Z, thereby enabling $\tilde{\xi}^{Zh}<1$. A measurement of $\tilde{\xi}^{Zh}<1$ would therefore constitute a clear signature of the NMSSM, as it depends on the singlet field and cannot be accomplished in the MSSM. In Fig.~\ref{fig:nmssm_ratio}, $\tilde{\xi}^{Zh}<1$ also occurs in the shaded region where $m_{\tilde\chi^0_2}-m_{\tilde\chi^0_1}<m_Z$ and $\tilde\chi^0_2$ decays only via three-body channels, which we do not consider as signal since as our 4$b$ on Z/h resonance signal vanishes. Note that a proper measurement of $\tilde{\xi}^{Zh}$ requires a lot of signal events, and therefore can only be accomplished at the high luminosity LHC; nevertheless, in the absence of other unambiguous signals, this can provide a method to distinguish the NMSSM from the MSSM. 

\section{Conclusions}
\label{sect:conclusion}

In this paper, we studied a very natural region of the NMSSM parameter space, favored by the 125 GeV Higgs and recent direct detection bounds, where Higgsino-like neutralinos can be directly produced at the 14 TeV LHC, and decay into the LSP and an on-shell Higgs. We considered several benchmark scenarios to study this process, studying the effects of different amounts of mixing with the singlet and the presence of a light pseudoscalar that can also be produced in the decay process.  

In particular, we studied the processes $\tilde\chi^0_{2,3}\rightarrow \tilde\chi^0_1+Z,h^*$ (where $h^*$ includes the observed Higgs and additional (pseudo)scalar close enough to 125 GeV to be indistinguishable in the decay process) with direct production of $\tilde\chi^0_{2,3}$ at the LHC, which contain little missing energy since $m_{\tilde\chi^0_2}-(m_h+m_{\tilde\chi^0_1})$ is small. Nevertheless, $pp\rightarrow \tilde\chi^0_{2,3}\tilde\chi^0_{2,3}jj$ offers a viable channel to look for these processes, and we estimate that realistic cuts can reduce the SM background to the same level as the signal. 

Correlating with spin-independent direct detection cross-sections, we found that the NMSSM can lead to cross-sections compatible with current bounds from LUX because of accidental numerical cancellations due to mixing in the Higgs sector, while the MSSM with positive $\mu$ leads to large cross sections and is unlikely to produce such signals at the LHC. We also looked at the relative ratio of $\tilde\chi^0_{2,3}$ decay into $Z$ and $h^*$, $\tilde{\xi}^{Zh}$ and found it to be in the range $1-2$ in most of the MSSM and NMSSM parameter space. An exception occurs when a (pseudo)scalar has mass very close to 125 GeV so that it is indistinguishable from the observed Higgs and emerges in the decays, and in such cases $\tilde{\xi}^{Zh}<1$ is possible, and constitutes a striking signature of the NMSSM. These can therefore act as useful handles to discriminate the underlying model if such a signal is observed at the LHC.
 
\section{Acknowledgements}

We thank Teruki Kamon,  Nikolay Kolev and Sean Wu  for helpful discussions and providing the $b$ faking efficiencies in estimating the $t\bar{t}$ background sample. The work of B.D. is supported by DOE Grant DE-FG02-13ER42020. Y.G. thanks the Mitchell Institute for Fundamental Physics and Astronomy for support. BS is supported by the DoE under grant DE-SC0007859. BS also thanks the Mitchell Institute for Fundamental Physics and Astronomy, where part of this work was completed, for hospitality.

\end{document}